\documentstyle[11pt,aaspp4]{article}

\received{1996 August 1}
\accepted{1996 September 12}
\journalid{337}{15 January 1989}
\articleid{11}{14}

\begin{document}     

\newcommand{\squig}{$\sim$}
\newcommand{\decsec}[2]{$#1\mbox{$''\mskip-7.6mu.\,$}#2$}
\newcommand{\decsectim}[2]{$#1\mbox{$^{\rm s}\mskip-6.3mu.\,$}#2$}
\newcommand{\decmin}[2]{$#1\mbox{$'\mskip-5.6mu.$}#2$}
\newcommand{\asecbyasec}[2]{#1$''\times$#2$''$}
\newcommand{\aminbyamin}[2]{#1$'\times$#2$'$}

\title{An Ultraviolet-Excess Optical Candidate for the
Luminous Globular Cluster X-ray Source in NGC\,1851
\footnote{\ Based on observations with the NASA/ESA Hubble Space
Telescope obtained at the Space Telescope Science Institute, which is
operated by the Association of Universities for Research in Astronomy,
Inc., under NASA contract NAS5-26555.}
}
\author{Eric W. Deutsch, Scott F. Anderson, and Bruce Margon}
\affil{Department of Astronomy, 
       University of Washington, Box 351580,
       Seattle, WA 98195-1580\\
       deutsch@astro.washington.edu; anderson@astro.washington.edu;
       margon@astro.washington.edu}

\author{and\\ \vskip .1in Ronald A. Downes}
\affil{Space Telescope Science Institute,
       3700 San Martin Drive,
       Baltimore, MD 21218\\
       downes@stsci.edu}

\begin{center}
Accepted for publication in The Astrophysical Journal Letters\\
To appear in volume 472, 1996 December 1\\
{\it received 1996 August 1; accepted 1996 September 19}
\end{center}

\begin{abstract}

The intense, bursting X-ray source in the globular cluster NGC\,1851
was one of the first cluster sources discovered, but has remained
optically unidentified for 25 years.  We report here on results from
{\it Hubble Space Telescope} WFPC2 multicolor images in NGC\,1851.  Our
high spatial resolution images resolve \squig 200 objects in the 3$''$
radius {\it Einstein} X-ray error circle, 40 times as many as in
previous ground-based work.  A color-magnitude diagram of the cluster
clearly reveals a markedly UV-excess object with B\squig 21, (U$-{\rm
B})\sim -0.9$ only 2$''$ from the X-ray position.

The UV-excess candidate is \decsec{0}{12} distant from a second,
unremarkable star that is 0.5 mag brighter in B; thus ground-based
studies of this field are probably impractical.  Three other UV-excess
objects are also present among the \squig 16,000 objects in the
surveyed region of the cluster, leaving a \squig 5\% probability that a
UV-excess object has fallen in the X-ray error circle by chance.  No
variability of the candidate is seen in these data, although a more
complete study is required.  If this object is in fact the counterpart
of the X-ray source, previous inferences that some globular cluster
X-ray sources are optically subluminous with respect to low-mass X-ray
binaries in the field are now strengthened.

\end{abstract}


\clearpage
\section{INTRODUCTION}

It has been known for over two decades that cluster X-ray sources are
overabundant with respect to those in the field (Katz 1975; Clark
1975).  While globular clusters only contain $10^{-4}$ of the mass of
the Galaxy, over $10^{-2}$ of the bright X-ray sources (L$_{\rm X}\geq$
$10^{36}$ erg s$^{-1}$) are found in globulars.  The cause of this
discrepancy may be a difference in the nature of globular cluster
X-ray sources from those in the field or more likely an enhancement of
their formation probability.

Despite a quarter-century of effort, there are very few intense globular
cluster X-ray sources with convincingly-identified optical
counterparts.  AC 211 in M\,15 (Auri\`ere et al. 1984; Charles et al.
1986) was discovered over a decade ago; it is surprisingly optically
luminous.  A variety of candidates in NGC\,6712 were offered (e.g.
Cudworth 1988; Nieto et al. 1990; Bailyn et al. 1991; Auri\`ere \&
Koch-Miramond 1992) and the counterpart, Star S, was eventually
unambiguously confirmed by Anderson et al. (1993) with {\it HST}
images.  A counterpart for the X-ray source in NGC\,6624 was discovered
to dominate the cluster light at 1400 \AA\ in {\it HST} FOC images by
King et al. (1993).  These three optical counterparts discovered so far
show a wide range in optical luminosity, and recently the spectra of
AC~211 and Star S were found to differ significantly (Downes et al.
1996).

The scarcity of optical counterparts is almost certainly primarily due
to the extreme crowding in the clusters, which limits the utility of
ground-based identification programs.  As part of the {\it Hubble Space
Telescope} Faint Object Spectrograph team's observing program, we have
obtained multicolor {\it HST} images of several globular clusters
containing X-ray sources, with the aim of resolving optical
counterparts in crowded fields and studying their spectra.  We report
here on the results from WFPC2 images in NGC\,1851.

X-ray observations of the strong source in NGC\,1851 are virtually as
old as satellite X-ray astronomy, dating back to the days of OSO--7
(Clark et al. 1975); Forman \& Jones (1976) first observed the bursting
nature of X0512--401 with {\it Uhuru}.  Recent X-ray observations of
the system have been made and reviewed by Callanan et al. (1995).  This
source would seem a particularly good candidate for an optical
identification, as there is an accurate {\it Einstein} HRI X-ray
position (Grindlay et al. 1984), and it lies over two core radii
(\squig 12$''$) north of the cluster center.

Ground-based imaging studies in NGC\,1851 by Auri\`ere et al. (1994;
hereafter ABK) do not reveal any especially good optical candidate
(e.g., no object with marked UV-excess) for the X-ray source
X0512--401.  However, these authors call attention to an object they
denote as X1, as the best candidate they are able to resolve in a 3$''$
error circle.  As its color (U$-$B$\sim$0) is no more UV than many hot
blue horizontal branch stars in the cluster, ABK speculate that X1 may
be a new sort of low-mass X-ray binary (LMXRB), in which there is very little
X-ray heating of the accretion disk.

Previous studies of NGC\,1851 using preservicing-mission {\it HST}
images did not detect any better candidates (Margon et al. 1992;
Deutsch et al.  1994).  The WF/PC images used in those studies do not
go nearly as deep as the WFPC2 data used in the present work.  Deutsch
et al. (1994) reported several objects which were bluer than X1 at the
detection threshold of those images, but were unable to isolate any
single object unusual enough to be considered a better candidate.
These earlier studies would have easily revealed counterparts as
luminous as AC 211 or Star~S.  Clearly, deeper exposures are required
to look for fainter objects in NGC\,1851.

\section{OBSERVATIONS AND DATA REDUCTION}

\subsection{Planetary Camera Imagery}

On 1996 April 10, we obtained {\it HST} WFPC2 images of NGC\,1851.  Two
900 s exposures through the F336W filter were taken, followed by four
300 s F439W exposures, and finally two more 900 s F336W exposures; the
F336W and F439W filters are similar to Johnson U and B respectively.
The frames have been processed through the standard data reduction
pipeline at STScI.  Further reduction was performed with software
written in IDL by E.W.D.  or available in the IDL {\it Astronomy User's
Library} (Landsman 1993).  Some photometry was performed with DoPHOT
(Schechter et al.  1993).

Although the images were obtained with the fine-lock mode of {\it HST},
the PSFs clearly show elongation in the Y direction in the Planetary
Camera (PC) images, with major to minor axis ratios of about 1.25.
Furthermore, while the images show excellent registration in the X
direction, Y offsets between exposures through the same filter are as
high as 0.4 pixels.

For each filter, the set of four images is combined into one master
image using software written by E.W.D.  First, cosmic-ray hits are
identified in each image and those pixels are set to zero in the data
and ``exposure mask" arrays.  The identification algorithm is similar
to the one decribed in Saha et al. (1996), and is carefully adjusted so
that the undersampled cores of bright stars with sub-pixel shifts
between images are not mistaken for cosmic-ray events.  The four images
and masks are aligned to \squig .01 pixel accuracy and summed.  The
summed frame is then divided by the mask to obtain the final image.  In
this way, cosmic-ray event pixels are excluded so that the information
in neighboring pixels is not contaminated during interpolation.

\subsection{Astrometry}

In order to determine the position of the X-ray source coordinates on
the {\it HST} PC image, we establish a coordinate system based on the {\it
HST} Guide Star Catalog (GSC) reference frame.  We begin with the
astrometric solution from a digitized Quick V image used to generate
the {\it HST} Guide Star Catalog (Lasker et al. 1990); this image,
obtained directly from ST\,ScI, contains the astrometric solution in
the image header.  By centroiding 29 stars in the Quick V image and the
corresponding objects in a short ground-based R-band CCD image, kindly
provided by H. Ford, the astrometric reference frame is transferred to
the CCD image with an error in the solution of \decsec{0}{1}.  Next,
the astrometric frame is transferred from the CCD image to the PC image
using a set of 9 common isolated stars, with error in the solution of
\decsec{0}{02}, negligible compared to the first step.  The final
result is that coordinates may be determined on the PC image to within
\decsec{0}{1} in the GSC reference frame.  However, there may well be some
systematic offset, $\sigma \sim$ \decsec{0}{5}, from frames based on
other reference catalogs (Russell et al. 1990).

We calculate the weighted mean of the X-ray coordinates from reprocessed {\it
Einstein} HRI data in the HRICFA database obtained through the High
Energy Astrophysics Science Archive Research Center Online Service,
provided by the NASA Goddard Space Flight Center.  These coordinates
are not significantly different from those in Grindlay et al. (1984),
so we use the published coordinates in this work.  The 90\% confidence
radius is reported by Grindlay et al. (1984) as 3$''$.

The optical position of the X-ray coordinates chosen by ABK, which we
estimate from Fig. 1 of that paper, is \squig 1$''$ northeast of our
location, which is not unreasonable given the different source of
astrometric reference.

The left panel of Fig. 1 shows a \asecbyasec{24}{24} region of the
cluster through the F336W filter.  A 3$''$ radius error circle is
overlaid, using our astrometric solution.  The right panel of Fig. 1
shows a \asecbyasec{8}{8} region of the cluster through the F336W
filter, centered on our optical position for the X-ray source, which we
have marked with a cross.

\subsection{Photometry}

Photometry was performed on the combined F336W and F439W PC frames with
the DoPHOT software (Schechter et al. 1993) and a PSF function-fitting
procedure written by E.W.D.  The results are similar, although DoPHOT
was not able to converge accurately on some of the brighter objects;
however, it performed better on the fainter objects.  For this reason,
we use the results derived from DoPHOT.  The fitted magnitudes were
calibrated with aperture photometry of several isolated objects.
Aperture corrections are taken from Table 2(a) in Holtzman et al.
(1995b).  The photometric measurements have not been corrected for
geometric distortions in the PC, but the simple correction for charge
transfer efficiency losses detailed in Holtzman et al. (1995b) has been
applied.  We use the photometric zero points for the STMAG system from
Table 9 in Holtzman et al. (1995a).  Systematic errors for all
magnitudes due to uncertainties in detector performance, absolute
calibration and filter transformations are \squig 5\%.

The photometric results are displayed in Fig. 2 as a color-magnitude
diagram of all \squig 5500 objects with reliable measurements in the PC frame.
Objects which are mentioned in the discussion below are marked in
the diagram.

\section{DISCUSSION}

While ABK were only able to resolve 5 stars within a 3$''$ radius error
circle, we are able to resolve and measure magnitudes for 194 objects
with $15\leq{\rm m}_{439}\leq 23$.

Star X1, suggested by ABK as the best candidate, does have the
brightest m$_{336}$ magnitude within the error circle.  However, its
color and magnitude are unremarkable compared with other cluster
horizontal branch stars.  We resolve the star labeled as X2 by ABK into
two main components, which we label X2a and X2b.  Additional, fainter
stars also contribute light to X2.  Object X2a is nearly in the center
of our error circle, but has normal colors.  In the CM diagram, X2b is
found to have a more UV color than star X1, but is also unremarkable
compared with other stars in the cluster (Fig. 2).

In the color-magnitude diagram of the entire \asecbyasec{34}{34} PC
field, there are two objects which stand out as having marked
UV-excess, with colors similar to known LMXRBs.  One of these is
\decsec{2}{15} away from our optical position for the X-ray
coordinates; this is well within the 90\% confidence error circle of
the X-ray coordinates.  We denote this object as star A and provide
magnitude measurements in Table~1.  Figure 3 shows a \asecbyasec{3}{3}
region of our PC data centered on Star A.  We also label Star B, which
is probably unrelated to, but blended with, Star A, at a separation of
only \decsec{0}{12}.  The photometry of these objects is slightly
hampered by the fact that the elongation of PSFs, discussed in \S 2.1,
is nearly in the same direction as the orientation of these two stars.
Future observations will be complicated by the proximity of the two
objects.

We have also examined the data for possible time variability of star
A.  The first and last F336W exposures are separated by 3 hr, while the
first and last F439W exposure are separated by only 70 min.  Neither
series shows any significant evidence for variability, although these
data do not allow an adequate study.  The 1$\sigma$ photometric
uncertainty from photon counting statistics is only 2\% and 3\% for
star A for the F336W and F439W filter exposures, respectively.
However, since Star A is blended with Star B, the measurements have a
higher uncertainty.  On the final F336W exposure, Star A was
contaminated with a cosmic-ray event, increasing the error of the
measurement for that exposure.  We set a variability upper limit of
0.10 mag during the time of our observations.  These data could not
have detected any small-amplitude, short-term variability of the scale
reported by Homer et al. (1996) for the globular cluster X-ray
source counterpart Star S in NGC\,6712.  Archival WF/PC images of NGC\,1851 have
also been examined for variability.  Star A is above the detection
threshold only in F336W exposures taken on 21 August 1993.  There is no
obvious evidence that Star A has changed in brightness during this 2.6
yr interval, but the variability upper limit is not very stringent,
\squig 0.4 mag.

Since star A is not the only UV-excess object in its part of the CM
diagram, there is some probability that it is not the X-ray
counterpart, but rather an unrelated object similar to the other
cluster UV objects.  To estimate the probability, we measure magnitudes
for all stars in the PC field as well as the three WF fields.  Although
this larger sample has $13\times$ more area, it contains only $3\times$
more stars than the PC field alone, as the latter is centered near the
cluster core, and there is a sharp radial gradient of stellar density.
Of the 16,000 stars for which we obtain good measurements, we find 4
objects with significant UV-excess of (m$_{336}-{\rm m}_{439})\leq
-0.5$ in the range $15\leq{\rm m}_{439}\leq 22$.  We also search each
of the fields with an image-subtraction technique to further insure
that no UV-excess objects are missed.

Therefore, approximately $2.5\times 10^{-4}$ stars with m$_{439}<22$ in
the cluster are markedly UV-excess, and as we measure 194 stars within
the error circle, we estimate a \squig 5\% probability that one of
these UV-excess objects unrelated to the X-ray source will be found in
the error circle by chance.  The hazards of these {\it a posteriori}
probability estimates are well-known, so this result must surely be
regarded as qualitative rather than quantitative.  The nature of the
faint UV-excess objects observed by {\it HST} in other globular cluster cores
has been recently discussed by numerous authors.

In NGC\,6712 the markedly UV-excess optical counterpart Star S is
surprisingly optically subluminous as compared with LMXRBs in the field.
(Anderson et al. 1993).  If the magnitude and color of Star S reported
in Homer et al. (1996) are corrected for the difference in reddening
and distance\footnote{A distance of 12.2 kpc and E(B$-$V$)=0.02$ for
NGC\,1851, and 6.8 kpc and E(B$-$V$)=0.46$ for NGC\,6712 were adopted
from Peterson (1993).  The reddening curve of Savage \& Mathis (1979)
was used.} between NGC\,6712 and NGC\,1851, one estimates that Star S,
were it present in NGC\,1851, would have m$_{439}=19.4$, (m$_{336}-{\rm
m}_{439})=-0.9$ in this CM diagram.  Therefore, while the colors of
Star S and the candidate presented here are similar, Star S is more
luminous by 1 mag.  If Star A is the correct identification of
the NGC\,1851 X-ray source, the luminosity discrepancy of this subclass
of objects from the field LMXRBs is further exacerbated.

To quantify this issue, we use the STSDAS {\it synphot} package, and find
that the blackbody spectrum which best fits the Star A color
(m$_{336}-{\rm m}_{439})=-0.73$ has T\squig 20,000~K.  Using a thermal
spectrum of this temperature, we derive approximate relations to
convert to Johnson U and B magnitudes: (U$-$m$_{336})=0.43$ and
(B$-$m$_{439})=0.65$.  Applying these zeropoint offsets, we estimate
U=20.2, B=21.1, (U$-{\rm B})=-0.9$.  A typical LMXRB in the field has
M$_{\rm V}\sim 1.2$, (B$-$V)$\sim 0$, and (U$-$B)$\sim
-0.9$ (van Paradijs 1983).  Therefore, the NGC\,1851 candidate has
(U$-$B) typical of LMXRBs, but is underluminous by over 4 mag, with
M$_{\rm B}=5.6$.  One model for the system described by Callanan et al.
(1995) invokes a subgiant secondary star; our faint counterpart rules
out this possibility.

\section{CONCLUSION}

We recompute and confirm the previously-published optical position of
the {\it Einstein} HRI $3''$ radius error circle for the bright X-ray
source in NGC\,1851 using the {\it HST} GSC as a reference frame.  With
WFPC2 observations 194 stars are resolved and photometered within the
error circle.  Star X1, the best candidate previously offered in the
literature, is resolved as a single star, but is not remarkable in a CM
diagram of the region.  Star X2 is resolved as two separate stars, the
brighter of which is more UV-excess than X1, but still not remarkable.

However, we find two UV-excess objects in the PC frame, one of which is
only \decsec{2}{15} from the X-ray source position.  Therefore, we
suggest this as a much more likely candidate for the optical
counterpart of the X-ray source, and we denote it as Star A.  We search
for, and find no variability for this object in these and archival
data.  However, the time coverage is poor, and a proper variability
study is desirable.  We find a total of 4 markedly UV-excess objects in
the \squig 16,000 objects measured in all four WFPC2 fields, and
therefore estimate that there is \squig 5\% probability that a
UV-excess star has fallen in the {\it Einstein} HRI $3''$ radius error
circle by chance.  Future studies of Star A will be complicated by the
presence of an unremarkable but comparably bright companion
\decsec{0}{12} distant.

Star S in NGC\,6712 is known to be significantly underluminous in the
optical compared to the typical LMXRBs in the field.  Star A in
NGC\,1851 is yet one magnitude fainter.  It will be interesting to
learn if we are merely sampling a broad underlying luminosity
dispersion, or are instead discerning a consequence of a physical
property of cluster LMXRBs that is fundamentally different from those
in the field.

\acknowledgments

We are grateful to Holland Ford for supplying a ground-based CCD image
of NGC\,1851, to Kerry McQuade for extracting the GSSS image from the
ST\,ScI archives, and to Ralph Bohlin, George Djorgovski and collaborators for
amicable sharing of {\it HST} data.  Financial support for this work
was provided by NASA through Grant NAG5-1630.


\clearpage

\begin{figure}
\plotone{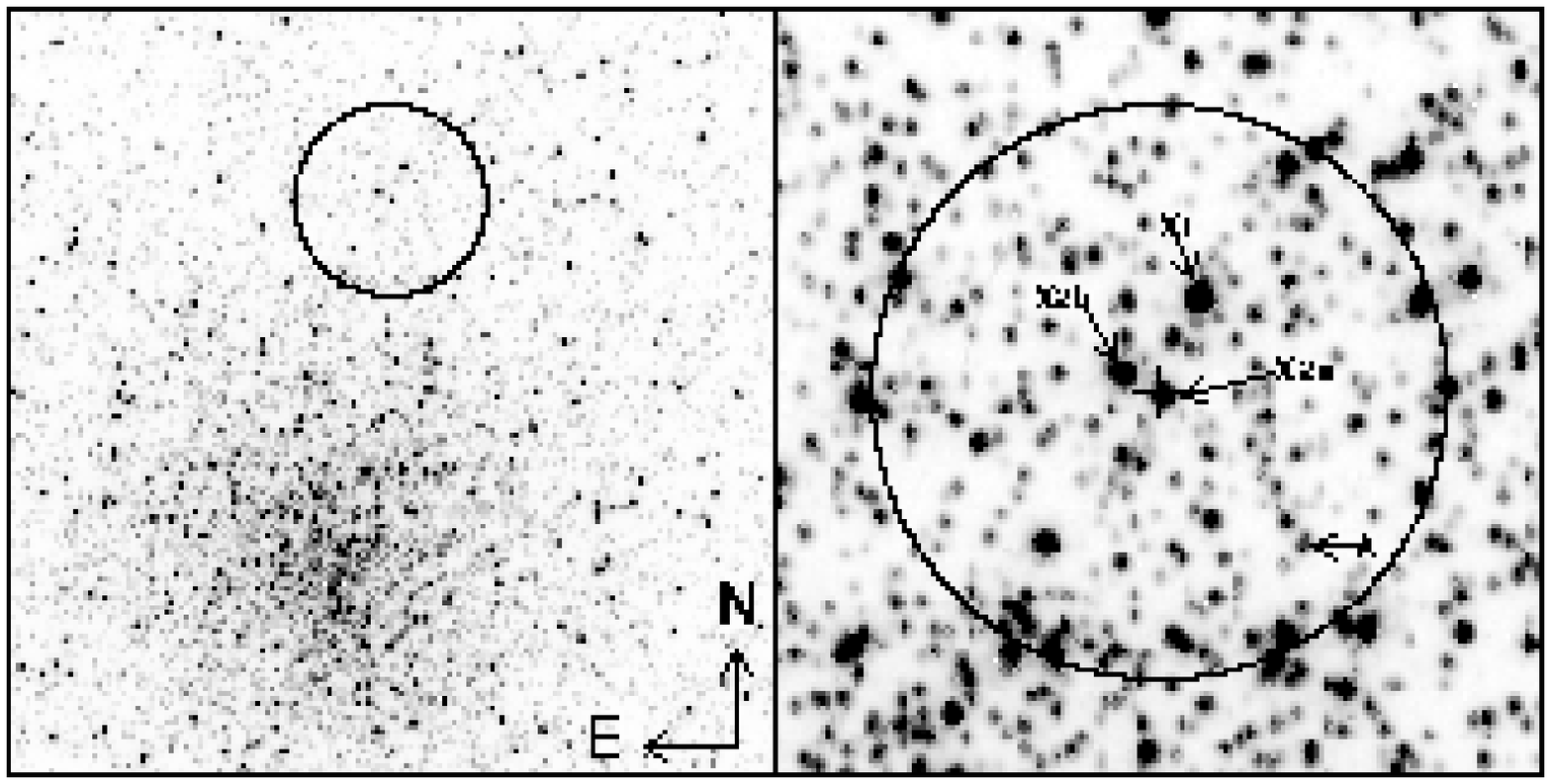}
\caption{{\it Left Panel}: \asecbyasec{24}{24} section of the {\it HST} PC
image of NGC\,1851 with the F336W filter.  The image has been
reoriented so that North is up and East is left.  A 3$''$ radius
error circle is drawn about the position of the X-ray coordinates.
{\it Right Panel}: \asecbyasec{8}{8} section of the same image centered
at the X-ray coordinates.  A cross denotes the optical position,
derived in this paper from the GSC source data, of the X-ray
coordinates, and the 3$''$ radius error circle is overlaid.  Several
objects discussed in the text are labeled.  Star A is selected as
the optical counterpart due to a significant UV excess.}
\end{figure}

\begin{figure}
\plotone{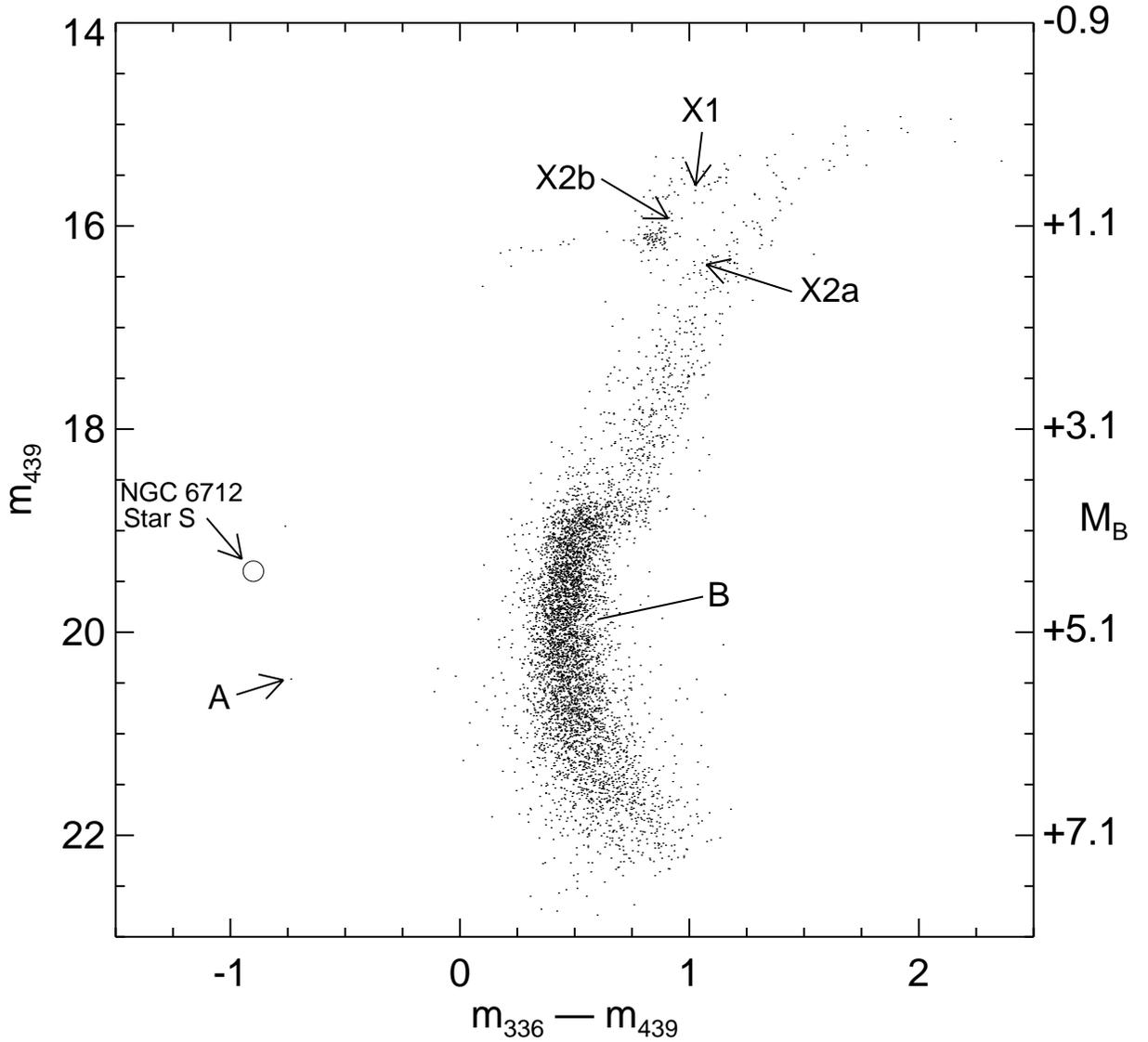}
\caption{A color-magnitude diagram for \squig 5500 stars in the PC
frame.  Magnitudes are in the STMAG system; see text for details.  All
objects which are discussed in the text are labeled here, including
Star A which we select as the optical counterpart due to a significant
UV excess.  An approximate M$_{\rm B}$ scale is also provided, assuming
(m$-$M)=15.43, E(B$-$V)=0.02, and (B$-$m$_{439})=0.65$.  This latter
correction changes slightly with stellar color; 0.65 is appropriate for
F type through the hottest stars, while 0.4 is more appropriate for M0
stars.  We also add the distance- and reddening-corrected magnitude and
color of Star S in NGC\,6712 for comparison.}
\end{figure}

\begin{figure}
\plotone{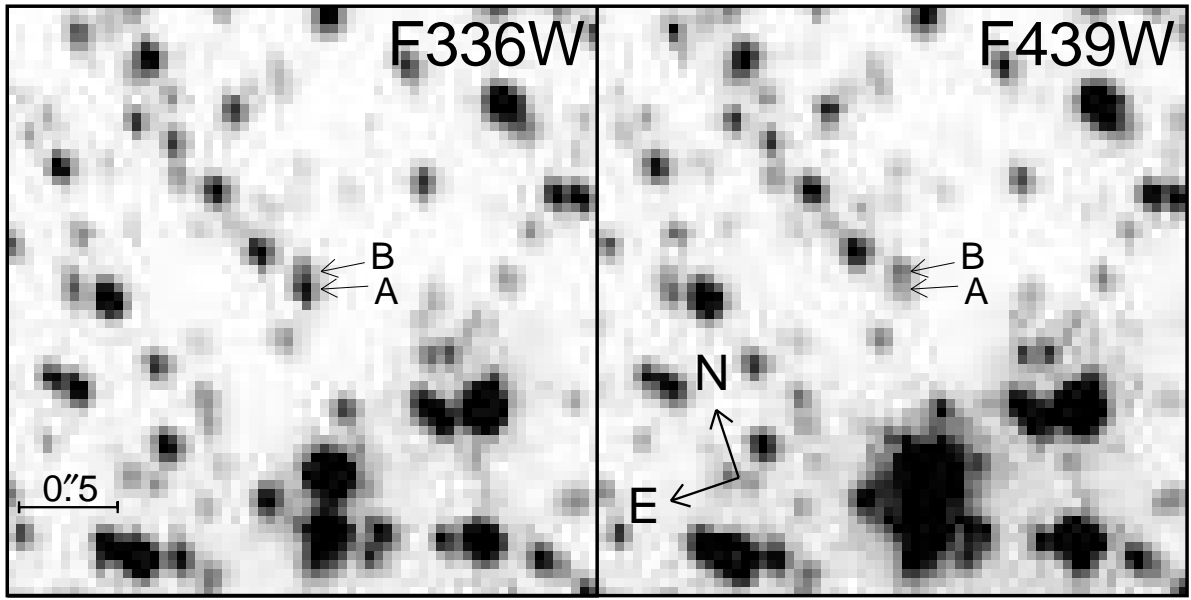}
\caption{\asecbyasec{3}{3} sections of the F336W (U) and F439W (B)
images centered on star A, which we select as the optical counterpart
due to a significant UV excess.  These images are still in the
original orientation, so North and East are labeled.
Star B, \decsec{0}{12} distant, is apparently unremarkable, but
may complicate future observations.}
\end{figure}

\clearpage
\begin{table}
Table 1.  Photometry and Coordinates for Selected Objects in NGC\,1851\\

\begin{tabular}{lccccccccc} \tableline \tableline
                            & $\alpha$(2000)  & $\delta$(2000)     \\
                            & 5${\rm ^h14^m}$ & $-40^\circ$02$'$   \\
Object                      &         (s)     &  (arcsec)   &  \multicolumn{1}{c}{m$_{336}$} & \multicolumn{1}{c}{m$_{439}$} &  (m$_{336}$--m$_{439})$  \\ \tableline
Star A   \tablenotemark{a}  &     6.428       &     38.38   &  19.73  &  20.46  & -0.73 \\
Star B                      &     6.425       &     38.26   &  20.43  &  19.92  &  0.51 \\
Star X1  \tablenotemark{b}  &     6.523       &     35.81   &  16.58  &  15.52  &  1.06 \\
Star X2a  \tablenotemark{c} &     6.556       &     36.81   &  17.51  &  16.40  &  1.11 \\
Star X2b  \tablenotemark{c} &     6.592       &     36.59   &  16.85  &  15.92  &  0.93 \\
\tableline
\end{tabular}
\tablenotetext{a}{ Optical counterpart for X0512$-$401 suggested in this work}
\tablenotetext{b}{ Designation from Auri\`ere et al. 1994}
\tablenotetext{c}{ X2, a designation from Auri\`ere et al. 1994, is multiply resolved in the current work}
\end{table}

\end{document}